\documentstyle[12pt]{article}

\begin{document}

\begin{titlepage}
\setcounter{page}{1}

\author{Waldemar Puszkarz\thanks{
Electronic address: puszkarz@cosm.sc.edu} \\
{\small {\it Department of Physics and Astronomy,} }\\
{\small {\it University of South Carolina,} }\\
{\small {\it Columbia, SC 29208}}}
\title{{\bf Higher Order Modification of the Schr\"{o}dinger Equation}}
\date{{\small (May 15, 1999)}}
\maketitle

\begin{abstract}
We modify the Schr\"{o}dinger equation in a way that preserves 
its main properties but makes use of higher order derivative terms. 
Although the modification represents an analogy to the Doebner-Goldin 
modification, it can differ from it quite distinctively. A particular model 
of this modification including derivatives up to the fourth order is 
examined in greater detail. We observe that a special variant of this model 
partially retains the linear superposition principle for the wave packets 
of standard quantum mechanics remain solutions to it. It is a peculiarity of
this variant that a periodic structure emerges naturally from its equations. 
As a result, a free particle, in addition to a plane wave solution, can 
possess band solutions. It is argued that this can give rise to well-focused
particle trajectories. Owing to this peculiarity, when interpreted outside 
quantum theory, the equations of this modification could also be used to 
model pattern formation phenomena. 

\vskip 0.5cm \noindent 
\end{abstract}

\end{titlepage}


Recently there has been a considerable interest in the Doebner-Goldin (DG)
modification of the Schr\"{o}dinger equation \cite{Doeb1}. (See also \cite
{Natt} for a more complete list of references to this subject and \cite
{Doeb2, Doeb3} for the latest update on the progress in a broader context
related to the matter in question.) It is the purpose of the present letter
to propose yet another modification of this fundamental equation that,
similarly as the Doebner-Goldin modification, makes use of the current
formulation and preserves the main features of the equation discussed, these
of homogeneity \cite{Gold1} and weak separability of composed systems \cite
{Bia}. Moreover, a subset of equations of this modification, defined by
certain values of its free parameters, complies with the Galilean
invariance, giving rise to a special version of it that, except for
linearity, possesses all the main properties of the Schr\"{o}dinger equation
deemed physically relevant.

Let us note that even though in some cases the homogeneity of nonlinear
generalizations of the Schr\"{o}dinger equation does entails their weak
separability \cite{Wein}, the properties mentioned are, in general,
independent \cite{Pusz1}. As pointed out in \cite{Pusz2}, the homogeneity of
nonlinear variants of this equation is essential for a unique defintion of
their energy functionals. This property is also necessary for modifications
of the Schr\"{o}dinger equation  to possess a weakly separable
multi-particle extension \cite{Pusz1, Pusz3}.

What differs our proposal from the DG modification is the use of terms that
involve derivatives of the order higher than second and of higher polynomial
degrees. As we will see, the simplest extension of this kind employes
derivatives of fourth order and of second degree. The higher degrees are
required for the completness of the formulation. As we will also see, a
certain, physically most attractive, variant of this modification should be
viewed as an extension of the Schr\"{o}dinger equation rather than its
modification for it changes the main properties of this equation less
dramatically than the majority of the modifications. In particular, this
true about its solutions.

To begin with, let us first reformulate the DG modification in a way that
will be convenient for the intended generalization. To this end, let us
write the modified Schr\"{o}dinger equation as 
\begin{equation}
i\hbar \frac{\partial \Psi }{\partial t}=\left( -\frac{\hbar ^{2}}{2m}\Delta
+V\right) \Psi -\frac{i\hbar D}{2}F_{\left\{ a\right\} }\left[ \Psi ,\Psi
^{*}\right] \Psi +\hbar DF_{\left\{ b\right\} }\left[ \Psi ,\Psi ^{*}\right]
\Psi ,  \label{1}
\end{equation}
where 
\[
F_{\left\{ x\right\} }\left[ \Psi ,\Psi ^{*}\right]
=\sum_{i=1}^{n}x_{i}F_{i}\left[ \Psi ,\Psi ^{*}\right] 
\]
and $x_{i}$ are some dimensionless coefficients that form a generic array $%
\left\{ x\right\} $ while $F_{i}\left[ \Psi ,\Psi ^{*}\right] $ are
functionals of $\Psi $ and $\Psi ^{*}$ homogeneous of degree zero in these
functions. The coupling constant $D$ has the dimensions of the diffusion
coefficient, meter$^{2}$second$^{-1}$, in the DG modification. In what
follows, we will work with $\rho =R^{2}=\Psi \Psi ^{*}$ and $S$, the
probability density and the phase of the wave function $\Psi =R\exp (iS)$,
correspondingly. The general form of the functional employed by Doebner and
Goldin is 
\begin{equation}
F_{\left\{ x\right\} }^{DG}\left[ \rho ,S\right] =x_{1}\Delta S+x_{2}\vec{%
\nabla}S\cdot \left( \frac{\vec{\nabla}\rho }{\rho }\right) +x_{3}\frac{%
\Delta \rho }{\rho }+x_{4}\left( \frac{\vec{\nabla}\rho }{\rho }\right)
^{2}+x_{5}\left( \vec{\nabla}S\right) ^{2}.  \label{2}
\end{equation}
The imaginary part of the Schr\"{o}dinger equation leads to the continuity
equation. The standard way to obtain it is to multiply both sides of the
Schr\"{o}dinger equation by $\Psi ^{*}$ and take the imaginary part of the
ensuing expression. The result turns out to be 
\begin{equation}
\frac{\partial \rho }{\partial t}+\frac{\hbar }{m}\vec{\nabla}\cdot \left(
\rho \vec{\nabla}S\right) =0  \label{3}
\end{equation}
with the probability current identified as $\vec{j}=\hbar \rho \vec{\nabla}%
S/m.$

In the same manner, we would like $\rho F_{\left\{ a\right\} }$ to form the
divergence of some current. One can show that two terms emerge to play this
role: $\vec{\nabla}\cdot \left( \rho \vec{\nabla}S\right) $ and $\Delta \rho 
$. One obtains these in a unique way by putting $a_{1}=a_{2}=a$ and $%
a_{4}=a_{5}=0$. Renaming $a_{3}D\rightarrow D^{\prime }$ and $aD\rightarrow
D $ allows us to write the modified continuity equation as\footnote{%
In the original DG formulation \cite{Doeb1}, $D$ does not make its
appearance at all. The case of nonzero $D$ is treated in a subsequent
generalization of their scheme involving nonlinear gauge transformations 
\cite{Doeb2}.} 
\begin{equation}
\frac{\partial \rho }{\partial t}+\frac{\hbar }{m}\vec{\nabla}\cdot \left(
\rho \vec{\nabla}S\right) +D\vec{\nabla}\cdot \left( \rho \vec{\nabla}%
S\right) +D^{\prime }\Delta \rho =0  \label{4}
\end{equation}
which can also be put in the form 
\begin{equation}
\frac{\partial \rho }{\partial t}+\frac{\hbar }{m^{*}}\vec{\nabla}\cdot
\left( \rho \vec{\nabla}S\right) +D^{\prime }\Delta \rho =0,  \label{5}
\end{equation}
where $m^{*}=m/\left( 1+Dm/\hbar \right) =m/\beta $ is the effective mass of
a quantum system. Let us notice that the most uniform way to write the last
equation is as follows: 
\begin{equation}
\frac{\partial \rho }{\partial t}+\frac{\hbar }{m^{*}}\vec{\nabla}\cdot
\left( \rho \vec{\nabla}S\right) +D^{\prime }\vec{\nabla}\cdot \left( \rho 
\frac{\vec{\nabla}\rho }{\rho }\right) =0.  \label{6}
\end{equation}
If we further note that $S=\frac{i}{2}\ln \left( \Psi ^{*}/\Psi \right) $
and $\vec{\nabla}\rho /\rho =\ln \Psi ^{*}\Psi $, we see that this way
reveals the ubiquitous role of the logarithmic function in the DG
modification\footnote{%
This should not come as a surprise as the logarithmic function was already
used for the sole purpose of ensuring that the modification of the
Schr\"{o}dinger equation proposed in \cite{Bia} obeys the weak separability
condition.} and suggests that the probability current be considered as
consisting of two components having velocities $\vec{v}_{S}=\frac{\hbar }{%
m^{*}}\vec{\nabla}S$ and $\vec{v}_{\rho }=D^{\prime }\vec{\nabla}\rho /\rho $%
. The latter is the diffusion component.

The linear Schr\"{o}dinger equation is invariant under the Galilean
transformation of coordinates $\vec{x}=\vec{x}^{^{\prime }}+\vec{v}%
t,t^{^{\prime }}=t$ provided the phase of the wave function transforms as 
\begin{equation}
S(\vec{x},t)=S^{^{\prime }}(\vec{x}^{^{\prime }},t^{^{\prime }})+m\vec{v}%
\cdot \vec{x}^{^{\prime }}+\frac{1}{2}m\vec{v}^{2}t^{^{\prime }}.  \label{7}
\end{equation}
One observes that the $D\vec{\nabla}\cdot \left( \rho \vec{\nabla}S\right) $
component of the additional current would break the Galilean invariance of
the continuity equation if it were not for the mass redefinition. To ensure
the Galilean invariance of this equation it is also required that the
effective mass replaces the ``bare'' mass everywhere a reference to the
latter is made, including in particular the phase of a quantum system in the
above transformational formula. However, the redefinition in question by no
means guarantees the Galilean invariance of the entire Schr\"{o}dinger
equation, which is maintained only if $D=0$.

It is also easy to notice that a special version of DG modification can be
straightforwardly derived by utilizing $\vec{\nabla}S$ and $\vec{\nabla}\rho
/\rho $ as the components of a general vector potential $A_{DG}=d_{1}\vec{%
\nabla}S+d_{2}\vec{\nabla}\rho /\rho .$ In the presence of $A_{DG}$ the free
Schr\"{o}dinger equation reads 
\begin{equation}
i\hbar \frac{\partial \Psi }{\partial t}=-\frac{\hbar ^{2}}{2m}\left( \vec{%
\nabla}-iA_{DG}\right) ^{2}\Psi .  \label{8}
\end{equation}
One arrives from it at the linearizable variant of the DG proposal that
generates the continuity equation in the form (4) or (5) and modifies the
other part of the Schr\"{o}dinger equation by adding terms $\left( \vec{%
\nabla}S\right) ^{2}$, $\left( \vec{\nabla}\rho /\rho \right) ^{2}$, and $%
\vec{\nabla}S\cdot \vec{\nabla}\rho /\rho $. The coefficients that stand by
these terms depend only on two constants which are functions of $d_{i}$. It
is clear now why this variant is linearizable and how to transform it into
the linear Schr\"{o}dinger equation. The latter is achieved by a nonlinear
gauge transformation \cite{Doeb2}, similarly as this is done in
electrodynamics. Some caution should be exercised when dealing with the $%
\vec{\nabla}S$ part of the gauge potential or the corresponding nonlinear
gauge transformation. As pointed out above this part is responsible for the
mass redefinition in the continuity equation. However, instead of the mass
redefinition, one can also enforce the Galilean invariance of this equation
by redefining the phase $S^{\prime }=\beta S$. Now, we see that this alters
the range of the phase from $2\pi $ to $2\pi \beta $ and should be taken
into account when dealing with the uniqueness of the wave function $\Psi $.%
\footnote{%
See also \cite{Czach1} for a more profound discussion of this issue in the
case the nonlinear gauge transformations are supposed to form a group.} Yet
another way to ensure the Galilean invariance is to redefine the Planck
constant, $\hbar \rightarrow \beta \hbar $, which eventually brings the
nonlinear Schr\"{o}dinger equation to the form of its linear prototype,
except that with the rescaled Planck constant. This completes the
linearization process. Being linearizable, the discussed variant of the DG
modification is physically equivalent to the linear Schr\"{o}dinger
equation, and, in particular, it satisfies the Ehrenfest equations. The DG
proposal in its full form is not linearizable. However, similarly as the
variant in question is related to the linear Schr\"{o}dinger equation via
nonlinear gauge transformations and thus constitutes a kind of family with
it, the equations of the full DG modification form a family that is closed
under more general gauge transformations \cite{Doeb2}.

No stipulation is made on the coefficients $\left\{ b\right\} $. However, if
one wants the modified Schr\"{o}dinger equation to be Galilean invariant,
(7) implies the unique choice of $b_{2}=b_{5}=0$. A restricted version of
the DG modification for $b_{2}=0$ and $2b_{3}-b_{4}=0$ in $F_{\left\{
b\right\} }\left[ \Psi ,\Psi ^{*}\right] $ was shown \cite{Pusz1} to be
derivable from a local Lagrangian density.

The main idea of our proposal is to extend the DG modification so as to
include higher derivative terms. Despite an apparent simplicity of this
task, the resulting construction turns out to be rather complex and can have
diametrically different features than the proposal of Doebner and Goldin. We
will first consider the leading order case, involving the derivatives up to
the fourth order and the second degree, which, in addition to its
simplicity, seems to be the most promising on physical grounds, and then
suggest how to generalize this construction to allow also higher order
terms. It is rather straightforward to convince oneself that 
\begin{eqnarray}
F_{\left\{ x\right\} }^{ext}\left[ \rho ,S\right] &=&x_{1}\Delta \Delta
S+x_{2}\Delta \left( \frac{\Delta \rho }{\rho }\right) +x_{3}\Delta \left( 
\frac{\vec{\nabla}\rho }{\rho }\right) ^{2}+x_{4}\Delta \left( \frac{\vec{%
\nabla}\rho }{\rho }\cdot \vec{\nabla}S\right) +  \nonumber \\
&&x_{5}\Delta \left( \vec{\nabla}S\right) ^{2}+x_{6}\frac{\vec{\nabla}\rho }{%
\rho }\cdot \vec{\nabla}\left( \Delta S\right) +x_{7}\frac{\vec{\nabla}\rho 
}{\rho }\cdot \vec{\nabla}\left( \frac{\Delta \rho }{\rho }\right) + 
\nonumber \\
&&x_{8}\frac{\vec{\nabla}\rho }{\rho }\cdot \vec{\nabla}\left( \frac{\vec{%
\nabla}\rho }{\rho }\right) ^{2}+x_{9}\frac{\vec{\nabla}\rho }{\rho }\cdot 
\vec{\nabla}\left( \frac{\vec{\nabla}\rho }{\rho }\cdot \vec{\nabla}S\right)
+x_{10}\frac{\vec{\nabla}\rho }{\rho }\cdot \vec{\nabla}\left( \vec{\nabla}%
S\right) ^{2}+  \nonumber \\
&&x_{11}\vec{\nabla}S\cdot \vec{\nabla}\Delta S+x_{12}\vec{\nabla}S\cdot 
\vec{\nabla}\left( \frac{\vec{\nabla}\rho }{\rho }\right) ^{2}+x_{13}\vec{%
\nabla}S\cdot \vec{\nabla}\left( \frac{\Delta \rho }{\rho }\right)  \label{9}
\end{eqnarray}
is the right choice for the homogeneous terms up to the fourth order. Terms $%
\left( \Delta S\right) ^{2}$, $\Delta S\left( \Delta \rho /\rho \right) $,
and $\Delta S\left( \vec{\nabla}\rho /\rho \right) ^{2}$ have not been
included. Even if homogeneous of degree zero in both $\Psi $ and $\Psi ^{*}$%
, they would nevertheless break the weak separability of the modified
Schr\"{o}dinger equation.

To cast more light on this issue, let us demonstrate the weak separability
of the Schr\"{o}dinger equation in the hydrodynamic formulation. We are
considering a quantum system made up of two noninteracting subsystems in the
sense that \cite{Bia} 
\begin{equation}
V(\vec{x}_{1},\vec{x}_{2},t)=V_{1}(\vec{x}_{1},t)+V_{2}(\vec{x}_{2},t).
\label{10}
\end{equation}
We will show that a solution of the Schr\"{o}dinger equation for this system
can be put in the form of the product of wave functions for individual
subsystems for any $t>0$, that is, $\Psi (x_{1},x_{2},t)=\Psi
_{1}(x_{1},t)\Psi _{2}(x_{2},t)=R_{1}(x_{1},t)R_{2}(x_{2},t)exp\left\{
i(S_{1}(x_{1},t)+S_{2}(x_{2},t))\right\} $ and that this form entails the
separability of the subsystems. The essential element here is that the
subsystems are initially uncorrelated which is expressed by the fact that
the total wave function is the product of $\Psi _{1}(\vec{x}_{1},t)$ and $%
\Psi _{2}(\vec{x}_{2},t)$ at $t=0$. What we will show then is that the
subsystems remain uncorrelated during the evolution and that, at the same
time, they also remain separated. It is the additive form of the total
potential that guarantees that no interaction between the subsystems occurs,
ensuring that they remain uncorrelated during the evolution. However, such
an interaction may, in principle, occur in nonlinear modifications of the
Schr\"{o}dinger equation even if the form of the potential itself does not
imply that. This is due to a coupling that a nonlinear term usually causes
between $\Psi _{1}(\vec{x}_{1},t)$ and $\Psi _{2}(\vec{x}_{2},t)$. As a
result, even in the absence of forces the very existence of one of the
particles affects the evolution of the other one, clearly violating
causality.

The discussed separability is called the weak separability since it assumes
that the wave function of the total system is the product of the wave
functions of its subsystems in contradistinction to the strong version of
separability that does not employ this assumption. As shown by L\"{u}cke 
\cite{Luc1, Luc2}, weakly separable modifications, such as the modification
of Bia\l ynicki-Birula \cite{Bia} or the Doebner-Goldin modification, \cite
{Doeb1} can still violate separability when the compound wave function is
not factorizable, and thus they are not strongly separable. An alternative
effective approach to the strong separability has been proposed by Czachor 
\cite{Czach2}. This approach treats the density matrix as the basic object
subjected to the quantum equations of motion which are modified\footnote{%
What this means in practice is that the basic equation is the nonlinear von
Neumann equation \cite{Czach3} instead of some nonlinear Schr\"{o}dinger
equation for the pure state.} compared to a nonlinear Schr\"{o}dinger
equation for the pure state. It admits a large class of nonlinear
modifications including those ruled out by the fundamentalist approach
advocated by L\"{u}cke and even those that are not weakly separable as, for
instance, the cubic nonlinear Schr\"{o}dinger equation.

The Schr\"{o}dinger equation for the total system, assuming that the
subsystems have the same mass $m$, reads now

\begin{eqnarray}
\lefteqn{\hbar \frac{\partial R_{1}^{2}R_{2}^{2}}{\partial t}+\frac{\hbar
^{2}}{m}\left\{ \left( \vec{\nabla}_{1}+\vec{\nabla}_{2}\right) \cdot \left[
R_{1}^{2}R_{2}^{2}\left( \vec{\nabla}_{1}S_{1}+\vec{\nabla}_{2}S_{2}\right)
\right] \right\} =\hbar R_{2}^{2}\frac{\partial R_{1}^{2}}{\partial t}+\hbar
R_{1}^{2}\frac{\partial R_{2}^{2}}{\partial t}}  \nonumber \\
&&+\frac{\hbar ^{2}}{m}R_{2}^{2}\vec{\nabla}_{1}\cdot \left( R_{1}^{2}\vec{%
\nabla}_{1}S_{1}\right) +\frac{\hbar ^{2}}{m}R_{1}^{2}\vec{\nabla}_{2}\cdot
\left( R_{2}^{2}\vec{\nabla}_{2}S_{2}\right) =R_{1}^{2}R_{2}^{2}\left\{
\left[ \hbar \frac{1}{R_{1}^{2}}\frac{\partial R_{1}^{2}}{\partial t}%
+\right. \right.  \nonumber \\
&&\left. \left. \frac{\hbar ^{2}}{m}\frac{1}{R_{1}^{2}}\vec{\nabla}_{1}\cdot
\left( R_{1}^{2}\vec{\nabla}_{1}S_{1}\right) \right] +\left[ \hbar \frac{1}{%
R_{2}^{2}}\frac{\partial R_{2}^{2}}{\partial t}++\frac{\hbar ^{2}}{m}\frac{1%
}{R_{2}^{2}}\vec{\nabla}_{2}\cdot \left( R_{2}^{2}\vec{\nabla}%
_{2}S_{2}\right) \right] \right\} =0  \label{11}
\end{eqnarray}

and

\begin{eqnarray}
\lefteqn{\frac{\hbar ^{2}}{m}\left( \Delta _{1}+\Delta _{2}\right)
R_{1}R_{2}-2\hbar R_{1}R_{2}\frac{\partial (S_{1}+S_{2})}{\partial t}-\frac{%
\hbar ^{2}}{m}R_{1}R_{2}\left( \vec{\nabla}_{1}S_{1}+\vec{\nabla}%
_{2}S_{2}\right) ^{2}-}  \nonumber \\
&&\left( V_{1}+V_{2}\right) R_{1}R_{2}=\frac{\hbar ^{2}}{m}R_{2}\Delta
_{1}R_{1}+\frac{\hbar ^{2}}{m}R_{1}\Delta _{2}R_{2}-2\hbar R_{1}R_{2}\frac{%
\partial S_{1}}{\partial t}-2\hbar R_{1}R_{2}\frac{\partial S_{2}}{\partial t%
}  \nonumber \\
&&+\frac{\hbar ^{2}}{m}R_{1}R_{2}\left( \vec{\nabla}_{1}S_{1}\right) ^{2}+%
\frac{\hbar ^{2}}{m}R_{1}R_{2}\left( \vec{\nabla}_{2}S_{2}\right)
^{2}-V_{1}R_{1}R_{2}-V_{2}R_{1}R_{2}=  \nonumber \\
&&R_{1}R_{2}\left\{ \left[ \frac{\hbar ^{2}}{m}\frac{\Delta _{1}R_{1}}{R_{1}}%
-2\hbar \frac{\partial S_{1}}{\partial t}+\frac{\hbar ^{2}}{m}\left( \vec{%
\nabla}_{1}S_{1}\right) ^{2}-V_{1}\right] +\left[ \frac{\hbar ^{2}}{m}\frac{%
\Delta _{2}R_{2}}{R_{2}}-2\hbar \frac{\partial S_{2}}{\partial t}+\right.
\right.  \nonumber \\
&&\left. \left. \frac{\hbar ^{2}}{m}\left( \vec{\nabla}_{2}S_{2}\right)
^{2}-V_{2}\right] \right\} =0,  \label{12}
\end{eqnarray}
where we used the fact that $\vec{\nabla}_{1}\cdot \vec{\nabla}_{2}=0.$ What
we have obtained is a system of two equations, each consisting of terms (in
square brackets) that pertain to only one of the subsystems. By dividing the
first equation by $R_{1}^{2}R_{2}^{2}$ and the second one by $R_{1}R_{2}$,
one completes the separation of the Schr\"{o}dinger equation for the
compound system into the equations for the subsystems. Moreover, we have
also showed that indeed the product of wave functions of the subsystems
evolves as the wave function of the total system. A similar analysis applied
to (9) convinces us that the chosen functional does have the required
property of weak separability. Of course, as already mentioned, this by no
means guarantees that the separability will be maintained for
nonfactorizable compound wave functions. However, the proposed modification
should be strongly separable in the effective approach \cite{Czach2}.

The coefficient $D$ has now the dimensions meter$^{4}$second$^{-1}$. By
choosing $a_{1}=a_{6}$, $a_{2}=a_{7}$, $a_{3}=a_{8}$, $a_{4}=a_{9}$, $%
a_{5}=a_{10}$, and $a_{11}=a_{12}=a_{13}=0$ one obtains the following
continuity equation 
\begin{eqnarray}
\frac{\partial \rho }{\partial t}+\frac{\hbar }{m}\vec{\nabla}\cdot \left(
\rho \vec{\nabla}S\right) +D_{1}\vec{\nabla}\cdot \left( \rho \vec{\nabla}%
\Delta S\right) +D_{2}\vec{\nabla}\cdot \left[ \rho \vec{\nabla}\left( \frac{%
\Delta \rho }{\rho }\right) \right] +  \nonumber \\
D_{3}\vec{\nabla}\cdot \left[ \rho \vec{\nabla}\left( \frac{\vec{\nabla}\rho 
}{\rho }\right) ^{2}\right] +D_{4}\vec{\nabla}\cdot \left[ \rho \vec{\nabla}%
\left( \frac{\vec{\nabla}\rho }{\rho }\cdot \vec{\nabla}S\right) \right]
+D_{5}\vec{\nabla}\cdot \left[ \rho \vec{\nabla}\left( \vec{\nabla}S\right)
^{2}\right] =0,  \label{13}
\end{eqnarray}
where $D_{i}=a_{i}D$. The new currents are revealed to be 
\begin{eqnarray}
\vec{j}_{1} &=&D_{1}\rho \vec{\nabla}\Delta S,\,\,\vec{j}_{2}=D_{2}\rho \vec{%
\nabla}\left( \frac{\Delta \rho }{\rho }\right) ,\,\,\vec{j}_{3}=D_{3}\rho 
\vec{\nabla}\left( \frac{\vec{\nabla}\rho }{\rho }\right) ^{2},\,\,\vec{j}%
_{4}=D_{4}\rho \vec{\nabla}\left( \frac{\vec{\nabla}\rho }{\rho }\cdot \vec{%
\nabla}S\right) ,  \nonumber \\
\vec{j}_{5} &=&D_{5}\rho \vec{\nabla}\left( \vec{\nabla}S\right) ^{2}.
\label{14}
\end{eqnarray}
The continuity equation (13) can also be written in the form 
\begin{equation}
\frac{\partial \rho }{\partial t}+\frac{\hbar }{m}\vec{\nabla}\cdot \left(
\rho \vec{\nabla}S\right) +D\vec{\nabla}\cdot \left[ \rho \vec{\nabla}%
F_{\left\{ a\right\} }^{DG}\left[ \rho ,S\right] \right] =0.  \label{15}
\end{equation}
Similarly as in the DG modification, no condition is put on the coefficients 
$\left\{ b\right\} $, but if one requires the modification to be Galilean
invariant one should demand that $%
b_{4}=b_{5}=b_{9}=b_{10}=b_{11}=b_{12}=b_{13}=0$. For the continuity
equation to be Galilean invariant one needs to set $D_{4}=D_{5}=0$.

One can ask if using $\vec{\nabla}F_{\left\{ a\right\} }^{DG}\left[ \rho
,S\right] $ as a vector potential in the linear Schr\"{o}dinger equation
would generate a special variant of the proposed modification similarly as
it was demonstrated for the linearizable part of the DG equations. It turns
out that this would not be so, which indicates that the modification
discussed may not be linearizable and therefore may contain some new physics
that cannot be described by the standard quantum theory. Such a formulation
would contribute the square of the vector potential to the real part of the
Schr\"{o}dinger equation. However, none of its terms, as for instance $%
\left( \vec{\nabla}\Delta S\right) ^{2}$, belongs to the modification under
study, being either of the higher order or degree in derivatives. Therefore,
to accomplish this, an extention of our proposal that would incorporate such
terms into it would be necessary. Even though this seems to be rather a
natural generalization of our work, it is not the purpose of the present
paper, and because of that it will not be discussed here.

It is not clear if there exists a local Lagrangian for the modification
proposed. It should be noted that not all equations of interest for
mathematical physics are derivable from local Lagrangians, the best case in
point being the celebrated Navier- Stokes equations with which no such
Lagrangian can be associated \cite{Fin}. It is also still unknown if the DG
modification in its fully developed form can be derived from any local
Lagrangian density.

An essential feature that differs this proposal from the DG modification is
the existence of unmodified stationary states of the Schr\"{o}dinger
equation, characterized by $S=-Et/\hbar $, for a certain set of parameters.
For this to occur, one needs to choose $D_{2}=D_{3}=0$ and $%
b_{2}=b_{3}=b_{7}=b_{8}=0$. If in addition to that we assume the Galilean
invariance, the Hamiltonian of the nonlinear Schr\"{o}dinger equation
reflecting these constraints reads 
\begin{equation}
H_{ME}=H_{L}+H_{NL}=H_{L}+\frac{i\hbar D_{1}}{2}\frac{\vec{\nabla}\cdot
\left( \rho \vec{\nabla}\Delta S\right) }{\rho }+\hbar b_{1}\Delta \Delta
S+\hbar b_{6}\frac{\vec{\nabla}\rho }{\rho }\cdot \vec{\nabla}\left( \Delta
S\right) ,  \label{16}
\end{equation}
where $H_{L}$ denotes the Hamiltonian of the linear Schr\"{o}dinger equation.%
\footnote{%
The subscript $ME$ stands for the {\bf minimal }higher order {\bf extension }%
of the Schr\"{o}dinger equation as explained further.} There may exist other
stationary states as well. The energy of stationary states is defined as the
expectation value of $H_{ME}$%
\begin{equation}
E_{ME}=\int dV\Psi ^{*}H_{ME}\Psi .  \label{17}
\end{equation}
It is straightforward to show that this definition leads to 
\begin{equation}
E_{ME}=E_{L}+\hbar \left( b_{1}-b_{6}\right) \int dV\rho \Delta \Delta S,
\label{18}
\end{equation}
where we assumed that the total current vanishes on a boundary in the
infinity and where $E_{L}$ is the expectation value of $H_{L}$ given by 
\begin{equation}
E_{L}=\int dV\left\{ \frac{\hbar ^{2}}{2m}\left[ \left( \vec{\nabla}R\right)
^{2}+\left( \vec{\nabla}S\right) ^{2}R^{2}\right] +VR^{2}\right\} .
\label{19}
\end{equation}
The general condition for stationary states, $\partial \rho /\partial t=0$,
leads to the equation 
\begin{equation}
\frac{\hbar }{m}\vec{\nabla}\cdot \left( \rho \vec{\nabla}S\right) +D_{1}%
\vec{\nabla}\cdot \left( \rho \vec{\nabla}\Delta S\right) =D_{1}\vec{\nabla}%
\cdot \left[ \rho \left( \vec{\nabla}\Delta S+\omega \vec{\nabla}S\right)
\right] =0,  \label{20}
\end{equation}
where $\omega =\frac{\hbar }{D_{1}m}$, which should be treated along with 
\begin{equation}
\frac{\hbar ^{2}}{m}\Delta \rho ^{1/2}-2\left( \hbar \frac{\partial S}{%
\partial t}+V+\frac{\hbar ^{2}}{2m}\left( \vec{\nabla}S\right) ^{2}+\frac{%
\hbar b_{1}}{2}\Delta \Delta S\right) \rho ^{1/2}-\hbar b_{6}\frac{\vec{%
\nabla}\rho }{\rho }\cdot \vec{\nabla}\left( \Delta S\right) \rho ^{1/2}=0.
\label{21}
\end{equation}
Together, they constitute a system of equations from which new stationary
states can be obtained. It is conceivable that these equations are satisfied
by wave functions whose phase $S$ depends also on the position. If $D_{1}>0$%
, one observes that even in the absence of the potential these equations
imply the existence of a periodic structure with a period $2\pi /\sqrt{%
\omega }$ that emerges naturally as a harmonic solution for $\vec{\nabla}S$.
It is a peculiarity of the discussed model that, unlike in some other
physical situations, this structure is not assumed but appears as a
consequence of the equations of motion. A free particle, in addition to a
plane wave solution, would display solutions characteristic of a particle
moving in a crystal, the band solutions. As observed in \cite{Car}, an
electron in a crystal is well focused in the same sense a particle in an
accelerator is. The equation that governs strong focusing in accelerators is
the same as the one that describes the motion of the electron in the
periodic crystalline lattice and entails the band structure of solids. It is
the Mathieu equation. In the discussed situation, one obtains this equation
from (21) by putting $b_{6}=0$. The strong focusing that this equation
implies may offer a novel solution to the problem of particle trajectory in
quantum mechanics. As is well known, in standard quantum mechanics the
particles are represented by wave packets. Yet, the wave packets in linear
theory spread without limit, and in so doing they contradict the existence
of sharp well defined particle trajectories that one observes in, say,
bubble chambers. In nonlinear quantum mechanics this problem can, in
principle, be solved. This is made possible by particle-like solutions, i.e., 
localized and non-spreading configurations known as solitons that are
generic for nonlinear equations of motion. Being well localized, the
solitons produce sharp trajectories in the bubble chamber. But, as we noted,
the equations of the modification under study can possibly give rise to
sharp well focused trajectories even if the particles are not solitons.
Interestingly, it is the phase of the wave function that entails this
remarkable solution, adding a new twist to the meaning of the wave-particle
duality.

On the other hand, if treated outside the quantum theory, the discussed
periodic solutions might provide a viable model for the pattern formation
phenomena \cite{Cro}. It should be noted in connection with this that the
equations of the Doebner-Goldin modification have their predecessor in a
phenomenological equation which originates in the study of the wave
propagation in fluids and plasmas with sharp boundaries and dissipation \cite
{Mal}.

As we see, with the assumptions of Galilean invariance and existence of
unmodified stationary states, the number of undetermined constants of our
modification reduces to only three, $D_{1}$ and $b_{1}$, and $b_{6}$. This
particular variant of the modification supports ordinary Gaussian wave
packets for which 
\begin{equation}
S=\frac{mtx^{2}}{2\left( t^{2}+t_{0}^{2}\right) }-\frac{1}{2}\arctan \frac{%
t_{0}}{t},  \label{22}
\end{equation}
(for simplicity in one dimension and in natural units) and the coherent
states satisfying $\Delta S=0$, and this again distinguishes it not only
from the DG modification that does not allow for any of these packets, but
also from the modifications of Staruszkiewicz \cite{Sta} and the one
proposed in \cite{Pusz1} which admit only the coherent states, excluding the
ordinary wave packets. The fact that the discussed variant admits ordinary
wave packets indicates that its nonlinearity is weak. It is a general
property of nonlinear modifications of the Schr\"{o}dinger equation to
exclude such packets. Since both ordinary Gaussian wave packets and coherent
states constitute the result of superposing more elementary wave functions,
each of which is a solution to the modification concerned, this means that
the variant in question maintains, if only partially, the linear
superposition principle. We choose to call this novel and rare property
among nonlinear modifications of the Schr\"{o}dinger equation the weak
nonlinearity. To the best of our knowledge, the property in question is
shared by only two other nonlinear modification of this equation \cite{Sta,
Pusz4}. We will call this variant of the modification the minimal higher
order extension of the Schr\"{o}dinger equation for it departs from this
equation in the most minimal way, preserving all of its standard properties,
including the stationary solutions.

One can easily show that the discussed version of the modification can also
be put in the form that involves a vector potential $\vec{A}=a\vec{\nabla}%
\Delta S$ as follows 
\begin{equation}
i\frac{\partial \Psi }{\partial t}=-\frac{\hbar }{2m}\left( \vec{\nabla}-i%
\vec{A}\right) ^{2}\Psi -\frac{\hbar }{2m}\vec{A}^{2}\Psi +\frac{c_{1}\hbar 
}{2m}\vec{\nabla}\cdot \vec{A}\Psi +\frac{\hbar }{2m}\left[ c_{2}Re(\vec{A}%
\cdot \vec{\nabla}\Psi )+2Im(\vec{A}\cdot \vec{\nabla}\Psi )\right] .
\label{23}
\end{equation}
The constant $a=-\frac{mD_{1}}{\hbar }$ has dimensions meter$^{2}$, the
other constants are defined as $c_{1}=\frac{2mb_{1}}{a}$ and $c_{2}=\frac{%
2mb_{6}}{a}$.

In general, the only notable exception to this rule being the modification
of Bia\l ynicki-Birula and Mycielski, nonlinear modifications of the
Schr\"{o}dinger equation do not have the classical limit in the sense of the
Ehrenfest theorem. The discussed variant of the modification is one of such
cases. The nonlinear terms it introduces entail corrections to the Ehrenfest
relations. We will now work out these corrections. For a general observable $%
A$ one finds that 
\begin{equation}
\frac{d}{dt}\left\langle A\right\rangle =\frac{d}{dt}\left\langle
A\right\rangle _{L}+\frac{d}{dt}\left\langle A\right\rangle _{NL},
\label{24}
\end{equation}
where the nonlinear contribution is due to $H_{NL}=H_{R}+iH_{I}$, $H_{R}$
and $H_{I}$ representing the real and imaginary part of $H_{NL}$,
respectively. The brackets $<>$ denote the mean value of the quantity
embraced. Specifying $A$ for the position and momentum operators, one
obtains the general form of the modified Ehrenfest relations \cite{Pusz1} 
\begin{equation}
m\frac{d}{dt}\left\langle \vec{r}\right\rangle =\left\langle \vec{p}%
\right\rangle +I_{1},  \label{25}
\end{equation}
\begin{equation}
\frac{d}{dt}\left\langle \vec{p}\right\rangle =-\left\langle \vec{\nabla}%
V\right\rangle +I_{2},  \label{26}
\end{equation}
where 
\begin{equation}
I_{1}=\frac{2m}{\hbar }\int dV\vec{r}\rho H_{I},  \label{27}
\end{equation}
\begin{equation}
I_{2}=\int dV\rho \left( 2H_{I}\vec{\nabla}S-\vec{\nabla}H_{R}\right) .
\label{28}
\end{equation}
In the derivation of the last formula it was assumed that $\int dV\vec{\nabla%
}\left( \rho H_{I}\right) =0$ which indeed stems from the continuity
equation. For the extension in question these integrals are found to be 
\begin{equation}
I_{1ME}=D_{1}m\int dV\vec{r}\vec{\nabla}\cdot \left( \rho \vec{\nabla}\Delta
S\right) ,  \label{29}
\end{equation}
\begin{equation}
I_{2ME}=\hbar \int dV\left[ D_{1}\vec{\nabla}\cdot \left( \rho \vec{\nabla}%
\Delta S\right) \vec{\nabla}S-b_{1}\rho \vec{\nabla}\Delta ^{2}S-b_{6}\rho 
\vec{\nabla}\left( \frac{\vec{\nabla}\rho }{\rho }\cdot \vec{\nabla}\Delta
S\right) \right] .  \label{30}
\end{equation}

One can straightforwardly generalize this construction to even higher order
derivatives. To this end, we note that the total current functional in (15)
is what was the functional from which to build the DG modification.
Similarly, the next order current functional will be the functional of
formula (9). The complexity of the higher order extensions makes them
difficult to study. On the other hand, this may even not be physically
justifiable or interesting. Equations involving fourth order derivatives do
occur in nonlinear equations aimed at modeling physical phenomena, such as,
for instance, the pattern formation \cite{Cro}. Higher order derivatives are
rather uncommon.

The question that deserves further study concerns the new currents. It is
quite natural to inquire if they form any algebraic structure.Of particular
interest is the issue of strong separability of the modification proposed in
the fundamentalist approach.

It would also be interesting to compare this modification with the one put
forward by Staruszkiewicz \cite{Sta} and extended by this author \cite{Pusz3}
which is neither homogeneous nor possesses the property of weak separability
and with the modification proposed in \cite{Pusz1} which is homogeneous but
does not automatically admit the weak separability of composed systems. The
common feature of these modifications is the presence of higher degrees of
derivatives or derivatives of the order higher than second. Similar
comparisons would not be possible for modifications with lower degrees of
derivatives. This provides a unique framework for a better understanding of
the concepts of weak separability and homogeneity, their physical impact and
possible connections.

The progress in these matters will be reported elsewhere.

\section*{Acknowledgments}

I would like to thank Professor Pawe{\l } O. Mazur for bringing to my
attention the work of Professor Staruszkiewicz which started my interest in
nonlinear modifications of the Schr\"{o}dinger equation and Professor Gerald
Goldin for his interest in this work. A stimulating exchange of
correspondence with Dr. Marek Czachor  and a correspondence with Professor 
Wolfgang L\"{u}cke on the problem of separability in nonlinear quantum 
mechanics are is also gratefully acknowledged. This work was partially 
supported by the NSF grant No. 13020 F167 and the ONR grant R\&T No. 3124141.

\end{document}